\documentclass{amsart}
\usepackage{amsmath,amssymb,amsfonts,epsfig}

\newtheorem{theorem}{Theorem}[section]

\theoremstyle{definition}

\theoremstyle{remark}

\numberwithin{equation}{section}

\def\DJ{{\hbox{D\kern-.8em\raise.15ex\hbox{--}\kern.35em}}}
\def\DJo{\DJ okovi\'c}

\providecommand{\bysame}{\leavevmode\hbox to3em{\hrulefill}\thinspace}

\font\germ=eufm10

\def\su{{\mbox{\germ su}}}

\def\vf{{\varphi}}
\def\bC{{\bf C}}

\def\bZ{{\bf Z}}

\def\pH{{\mathcal H}}

\def\pP{{\mathcal P}}

\def\tr{{\rm \;tr}}

\def\SU{{\mbox{\rm SU}}}

\let\la=\langle  \let\ra=\rangle

\begin{document}

\title[ Poincar\'{e} series for the qubit-qutrit system ]
{Poincar\'{e} series for local unitary invariants
of mixed states of the qubit-qutrit system}

\author[D.\v{Z}. \DJo{}]
{Dragomir \v{Z}. \DJo{}}

\address{Department of Pure Mathematics, University of Waterloo,
Waterloo, Ontario, N2L 3G1, Canada}

\email{djokovic@uwaterloo.ca}

% \subjclass{Primary 13A50, 68W30; Secondary 13P10}

\thanks{The author was supported in part by the NSERC
Grant A-5285.}

\date{}

\begin{abstract}
We consider the mixed states of the bipartite quantum system
with the first party a qubit and the second a qutrit.
The Hilbert space of this system is the tensor product
$\pH=\bC^2\otimes\bC^3$ and the group of local unitary 
transformations, ignoring the overall phase factor, is the
direct product $G=\SU(2)\times\SU(3)$.
Let $\pP$ be the algebra of real polynomial functions on the
affine space of all hermitian operators of trace 1 on $\pH$.
The polynomials $f\in\pP$ which are invariant under
$G$ form the subalgebra $\pP^G\subseteq\pP$.
We compute the simply graded Poincar\'{e} series of this
subalgebra and construct several low degree invariants.
\end{abstract}

\maketitle

\section{Introduction}

Let $\bC^n$ be the Hilbert space (of column vectors)
with the inner product $\la x_1,x_2 \ra=x_1^\dag x_2$
and $M_n$ the algebra of complex $n\times n$ matrices.
For $X,Y\in M_n$, their inner product is
defined by $\la X_1,X_2 \ra=\tr(X_1^\dag X_2)$.
Let $\su(n)$ denote the Lie algebra of $\SU(n)$, it consists
of all traceless skew-hermitian matrices in $M_n$.

We consider the bipartite quantum system with the first
party a qubit, $\bC^2$, and the second one a qutrit, $\bC^3$.
It will be convenient to identify $\bC^6$ with the tensor product
$\pH=\bC^2\otimes_\bC \bC^3$ and $M_6$ with $M_2\otimes_\bC M_3$.
For $X_1,X_2\in M_2$ and $Y_1,Y_2\in M_3$, the inner product of
$X_1\otimes Y_1$ and $X_2\otimes Y_2$ is given by
\[ \la X_1\otimes Y_1,X_2\otimes Y_2 \ra=
\la X_1,X_2 \ra\cdot \la Y_1,Y_2 \ra. \]
It follows that $(X\otimes Y)^\dag=X^\dag\otimes Y^\dag$ and,
in particular, the tensor product of hermitian matrices is
again a hermitian matrix.

The space of hermitian matrices $H_n\subseteq M_n$ is the direct
sum of the $1D$ real space spanned by the identity matrix, $I_n$,
and the space $H_{n,0}=i\su(n)$. The space
$H_{6,0}$ is the direct sum of three real subspaces:
$V_1=H_{2,0}\otimes I_3$, $V_2=I_2\otimes H_{3,0}$,
and $V_3=H_{2,0}\otimes H_{3,0}$.
Any mixed state, $\rho$, of this quantum system can be written
uniquely as the sum of four components;
\begin{equation} \label{stanje}
\rho=\frac{1}{6}I_6+X\otimes I_3+I_2\otimes Y+Z,
\end{equation}
where $X\in H_{2,0}$, $Y\in H_{3,0}$, and $Z\in V_3$. If we
fix a basis $\{E_1,E_2,E_3\}$ of $H_{2,0}$, then
$Z$ can be written uniquely as
\begin{equation} \label{vektor}
Z=\sum_{k=1}^3 E_k\otimes Y_k,\quad Y_k\in H_{3,0}.
\end{equation}
For instance, we could take the basis consisting of the three
Pauli matrices.

Denote by $G$ the group of local unitary transformations,
$\SU(2)\times\SU(3)$, where we ignore the overall phase factor.
Note that $G$ acts on $M_6$ in the usual manner:
$(g,Z)\to gZg^\dag$. It stabilizes the real subspaces
$H_6$ and $H_{6,0}=V_1+V_2+V_3$.
Moreover, each of the subspaces
$V_1$, $V_2$, $V_3$ is a simple $G$-module.

Let $\pP$ denote the algebra of real valued polynomial functions
on $H_{6,0}$ and $\pP^G$ the subalgebra of $G$-invariant functions.
Note that $\pP^G$ inherits the $\bZ$-gradation from $\pP$.
The homogeneous polynomials of $\pP^G$ may be used to construct
measures of entanglement of this quantum system.

The main objective of this note is to determine the Poincar\'{e}
series (also known as the Hilbert series) of $\pP^G$.
The knowledge of the algebra $\pP^G$ is important because of
the following well known fact: Two states, say $\rho_1$ and
$\rho_2$, belong to different $G$-orbits iff
there exists $f\in\pP^G$ such that $f(\rho_1)\ne f(\rho_2)$.
Due to the direct decomposition of $H_{6,0}$ into three $G$-invariant
subspaces, $\pP$ also admits a $G$-invariant $\bZ^3$-gradation.
It induces $\bZ^3$-gradation on $\pP^G$, to which we refer as
the multigradation. So far we did not succeed to compute
the multigraded Poincar\'{e} series of $\pP^G$.

The case of two qubits has been considered by M. Grassl et al.
\cite{GRB}, Y. Makhlin \cite{Ma}, and also by the author \cite{DZ}.
The local invariants for pure states of more qubits have been
considered in several recent papers:
see \cite{AA,CS,LTT,MW,AS} for 3 qubits,
\cite{BLT,LT1,LTT} for 4 qubits, and \cite{LT2} for 5 qubits.

\section{The Poincar\'{e} series for the $2\otimes3$ case}

Going back to our case, the Poncar\'{e} series $P(t)$ is given by
\[ P(t)=\sum_{d=0}^\infty \dim(\pP_d^G) t^d, \]
where $\pP_d^G$ is the space of homogeneous polynomial
invariants of degree $d$. This Poincar\'{e} series is a rational
function of the variable $t$. It can be
computed by using the well-known Molien--Weyl formula.
Following the recipe from \cite{DK}, we obtain that
\[ P(t)=\frac{1}{(2\pi i)^3}
\int_{|z|=1} \int_{|y|=1} \int_{|x|=1} \vf(x,y,z,t)
\frac{ {\rm d}x }{x} \frac{ {\rm d}y }{y} \frac{ {\rm d}z }{z}, \]
where
\[ \vf(x,y,z,t)=\frac{(1-x^{-1})(1-y^{-1})(1-z^{-1})(1-y^{-1}z^{-1})}
{\psi(x,y,z,t)} \]
and
\begin{eqnarray*}
\psi &=& (1-t)^5(1-tx)^3(1-tx^{-1})^3(1-ty)^2(1-tz)^2(1-tyz)^2 \\
&& (1-ty^{-1})^2(1-tz^{-1})^2(1-ty^{-1}z^{-1})^2 (1-txy)(1-txz)(1-txyz) \\
&& (1-txy^{-1})(1-txz^{-1})(1-txy^{-1}z^{-1})
(1-tx^{-1}y)(1-tx^{-1}z)(1-tx^{-1}yz) \\
&& (1-tx^{-1}y^{-1})(1-tx^{-1}z^{-1})(1-tx^{-1}y^{-1}z^{-1}).
\end{eqnarray*}
The integrations are over the unit circle in the counterclockwise
direction, and one should assume that $|t|<1$.
The computation of these integrals was performed by using Maple \cite{Map}.
Our main result is the following.

\begin{theorem} \label{Poenk-red}
The Poincar\'{e} series of $\pP^G$ is the rational function
$P(t)$ whose numerator $N(t)$ and denominator $D(t)$ are given by
\begin{eqnarray*}
N &=& 1+t-2\,{t}^{3}+2\,{t}^{4}+13\,{t}^{5}+50\,{t}^{6}+102\,{t}^{7}
+216\,{t}^{8}+422\,{t}^{9}+874\,{t}^{10} \\
&& +1691\,{t}^{11}+3305\,{t}^{12}+6037\,{t}^{13}+10779\,{t}^{14}
+18312\,{t}^{15}+30318\,{t}^{16} \\
&& +48209\,{t}^{17}+74858\,{t}^{18}+112294\,{t}^{19}+164391\,{t}^{20}
+233394\,{t}^{21}+323332\,{t}^{22} \\
&& +435113\,{t}^{23}+571671\,{t}^{24}+730844\,{t}^{25}+912641\,{t}^{26}
+1110648\,{t}^{27} \\
&& +1321048\,{t}^{28}+1532768\,{t}^{29}+1739258\,{t}^{30}+1926469\,{t}^{31}
+2087251\,{t}^{32} \\
&& +2208470\,{t}^{33}+2286037\,{t}^{34}+2311126\,{t}^{35}+2286037\,{t}^{36}
+ \cdots \\
&& +2\,{t}^{66}-2\,{t}^{67}+{t}^{69}+{t}^{70}, \\
D &=&
 \left( 1+t \right)  \left( 1-{t}^{2} \right) ^{3} \left( 1-{t}^{3}
 \right) ^{6} \left( 1-{t}^{4} \right) ^{5} \left( 1-{t}^{5}\right) ^{4}
 \left( 1-{t}^{6} \right) ^{3} \left( 1-{t}^{7}
 \right) ^{2} \left( 1-{t}^{8} \right).
\end{eqnarray*}
\end{theorem}
(As $N$ is a palindromic polynomial of degree 70, it suffices to
write its coefficients up to degree 35.)

The Taylor expansion of $P(t)$ begins with the following terms
\begin{eqnarray*}
P(t) &=&
1+3\,{t}^{2}+4\,{t}^{3}+15\,{t}^{4}+25\,{t}^{5}+90\,{t}^{6}+170\,{t}^
{7}+489\,{t}^{8}+1059\,{t}^{9} \\
&& +2600\,{t}^{10}+5641\,{t}^{11}+12872\,{t}^{12}+27099\,{t}^{13}
+57990\,{t}^{14}+118254\,{t}^{15} \\
&& +240187\,{t}^{16}+472273\,{t}^{17}+919432\,{t}^{18}
+1745295\,{t}^{19}+ \cdots
\end{eqnarray*}

The numerator $N$ and the denominator $D$ are relatively prime.
After multiplying them with
$(1-t+t^2)(1+t^3)$, we obtain $P(t)=N^*(t)/D^*(t)$ where
$N^*$ is a palindromic polynomial with nonnegative integer coefficients
\begin{eqnarray*}
N^* &=& 1+4\,{t}^{4}+9\,{t}^{5}+38\,{t}^{6}+69\,{t}^{7}+173\,{t}^{8}
+347\,{t}^{9}+733\,{t}^{10}+1403\,{t}^{11} \\
&& +2796\,{t}^{12}+5091\,{t}^{13}+9286\,{t}^{14}+16058\,{t}^{15}
+27208\,{t}^{16}+44250\,{t}^{17} \\
&& +70537\,{t}^{18}+108430\,{t}^{19}+163158\,{t}^{20}+238264\,{t}^{21}
+339974\,{t}^{22} \\
&& +472130\,{t}^{23}+641187\,{t}^{24}+848615\,{t}^{25}+1098643\,{t}^{26}
+1388741\,{t}^{27} \\
&& +1717327\,{t}^{28}+2075836\,{t}^{29}+2456389\,{t}^{30}
+2843020\,{t}^{31}+3222408\,{t}^{32} \\
&& +3575226\,{t}^{33}+3884797\,{t}^{34}+4133599\,{t}^{35}+4308636\,{t}^{36}
+4398377\,{t}^{37} \\
&& +4398377\,{t}^{38}+ \cdots +38\,{t}^{69}+9\,{t}^{70}+4\,{t}^{71}+{t}^{75}
\end{eqnarray*}
and
\begin{equation*}
D^* = \left( 1-{t}^{2} \right) ^{3} \left( 1-{t}^{3}\right) ^{4}
\left( 1-{t}^{4} \right) ^{5} \left( 1-{t}^{5}\right) ^{4}
 \left( 1-{t}^{6} \right) ^{5} \left( 1-{t}^{7}
 \right) ^{2} \left( 1-{t}^{8} \right).
\end{equation*}

As the space $H_{6,0}\subseteq M_6$ of traceless hermitian matrices
has dimension 35 and the generic orbits of $G$ have dimension 11,
the orbit space of (mixed) states of the system has dimension 24.
Hence, the maximum number of algebraically independent generators of
$\pP^G$ is 24. On the basis of the expression $P(t)=N^*(t)/D^*(t)$
and the factorization of $D^*$,
it is quite reasonable to expect that a maximal set of algebraically
independent homogeneous polynomials in $\pP^G$ (in fact a homogeneous
system of parameters) can be chosen so
that three of them have degree 2, four degree 3, five degree 4,
four degree 5, five degree 6, two degree 7, and one degree 8.

\section{A few low degree invariants}

We assume in this section that the density matrix $\rho$ is
given by Eq. (\ref{stanje}) and that the component $Z$
is given by Eq. (\ref{vektor}).

From the above Taylor expansion of $P(t)$ we see that the space
of quadratic invariants is 3-dimensional.
The three linearly independent quadratic invariants are:
\begin{eqnarray*}
I_1(\rho) &=& \det(X), \\
I_2(\rho) &=& \tr(Y^2), \\
I_3(\rho) &=&  \tr(Z^2) = \sum_{k,l=1}^3 \tr(E_k E_l)\tr(Y_kY_l).
\end{eqnarray*}

The space of cubic invariants has dimension 4.
The four linearly independent cubic invariants are given by
\begin{eqnarray*}
I_4(\rho) &=& \det(Y), \\
I_5(\rho) &=& \tr(Z^3) = \sum_{k,l,m=1}^3 \tr(E_k E_l E_m)\tr(Y_kY_lY_m), \\
I_6(\rho) &=& \tr((X\otimes Y)Z) = \sum_{k=1}^3 \tr(XE_k) \tr(YY_k), \\
I_7(\rho) &=& \tr((I_2\otimes Y)Z^2)
= \sum_{k,l=1}^3 \tr(E_k E_l) \tr(YY_k Y_l).
\end{eqnarray*}
Of course, one has to check that each of these invariants is
nonzero. Then the linear independence follows from the fact
that they have different multidegrees, namely
$(0,3,0)$, $(0,0,3)$, $(1,1,1)$ and $(0,1,2)$, respectively.
(The first degree corresponds to the coordinates of $V_1$,
the second to $V_2$ and the third to $V_3$.)

\end{document}